\documentclass[Journal]{IEEEtran}
\IEEEoverridecommandlockouts
\usepackage{cite}
\usepackage{float}
\usepackage{amsmath,amssymb,amsfonts}
\usepackage[ruled,linesnumbered,vlined]{algorithm2e}
\usepackage{algpseudocode}
\usepackage{graphicx}
\usepackage{textcomp}
\usepackage{xcolor}
\usepackage{multirow}
\usepackage{diagbox}
\usepackage{array}
\usepackage{footnote}
\usepackage{makecell} 
\usepackage{footmisc}
\makesavenoteenv{tabular}
\makesavenoteenv{table}
\usepackage{subfigure}
\usepackage{lettrine}
\usepackage{hyperref}
\hypersetup{colorlinks,
citecolor=blue,
linkcolor=blue,
urlcolor=black}
\usepackage{pifont}

\begin{document}

\title{Resource and Mobility Management in Hybrid LiFi and WiFi Networks: A User-Centric Learning Approach}

\author{Han Ji, \textit{Student Member, IEEE}, Xiping Wu, \textit{Senior Member, IEEE}

\thanks{The work of H. Ji is supported by the China Scholarship Council (Grant No. 202106620012). X. Wu gracefully acknowledges the support of the National Natural Science Foundation of China (NSFC). This publication has emanated from research supported (in part) by the UCD STEM Challenge Fund. {\textit{Corresponding author: Xiping Wu}}}

\thanks{H. Ji is with the School of Electrical and Electronic Engineering, University College Dublin, Dublin, D04 V1W8, Ireland (e-mail: han.ji@ucdconnect.ie).}

\thanks{X. Wu is with the School of Information Science and Engineering, Southeast University, Nanjing 210096, China. He is also a visiting professor with University College Dublin, Dublin, D04 V1W8, Ireland (e-mail: xiping.wu@seu.edu.cn).}

}

\maketitle
\begin{abstract}
Hybrid light fidelity (LiFi) and wireless fidelity (WiFi) networks (HLWNets) are an emerging indoor wireless communication paradigm, which combines the complementary advantages of LiFi and WiFi. Meanwhile, load balancing (LB) becomes an essential and critical challenge, due to the nature of hybrid networks. The existing LB methods are mostly network-centric, relying on a central unit to make a solution for the users all at once. Consequently, the solution needs to be updated for all users at the same pace, regardless of their moving status. This would affect the network performance in two aspects: i) a lower update frequency would compromise the connectivity of fast-moving users; ii) a higher update frequency would cause unnecessary handovers as well as hefty feedback costs for slow-moving users. Motivated by this, we investigate user-centric LB so that users can update their solutions at different paces. The research is developed upon our previous work on adaptive target-condition neural network (ATCNN), which carries out LB for individual users in quasi-static channels. In this paper, a deep neural network (DNN) model is designed to enable an adaptive update interval for each individual user. This new model is termed as mobility-supporting neural network (MSNN). Associating MSNN with ATCNN, a user-centric LB framework named mobility-supporting ATCNN (MS-ATCNN) is proposed to handle resource management and mobility management simultaneously. Results show that at the same level of average update interval, MS-ATCNN can achieve a network throughput up to 215\% higher than conventional LB methods such as game theory (GT), especially for a larger number of users. In addition, MS-ATCNN costs an ultra-low inference time in sub-milliseconds, which is two to three orders of magnitude lower than the GT baseline.

\end{abstract}


\section{Introduction}

\lettrine[loversize=0.1, nindent=0em]{O}{N} the road to the sixth generation (6G) communication systems, hybrid light fidelity (LiFi) and wireless fidelity (WiFi) networks (HLWNets) are an emerging
paradigm of indoor wireless technologies \cite{wang20236G_road_Review}. LiFi operates in a way similar to WiFi but on the vast visible light spectrum (400-790 THz), offering remarkable advantages including license-free, electromagnetic interference-free, high physical layer security, etc \cite{haas2015lifi}. Recent experimental work shows that LiFi can provide a link data rate of 24 Gbps over a single light-emitting diode chip \cite{hu2021si}. In addition, LiFi can be integrated into the existing lighting infrastructure and provide illumination and communication simultaneously, rendering energy-saving potentials. Meanwhile, LiFi faces the challenges of light-path blockage caused by opaque objects such as human bodies and furniture \cite{wu2023blockage}, despite the non-line-of-sight (NLoS) paths can facilitate the data transmission to some extent \cite{zhu2020indoorNLoS}. Also, LiFi offers a relatively small coverage area with a single access point (AP), usually around a few metres in diameter. In contrast, the WiFi AP gives a larger coverage range (up to 50 meters) but a lower throughput, which achieved 92 Mbps in average by 2023 \cite{cisco2020cisco}. The complementary advantages of LiFi and WiFi motivate the co-existence of them in the indoor wireless systems, composing HLWNets which have gained increasing attention in recent years \cite{wu2021hybrid}. Such hybrid networks can significantly improve the network capacity over stand-alone LiFi or WiFi networks \cite{basnayaka2015hybrid}.

In HLWNets, the coverage of LiFi and WiFi APs overlaps each other, imposing a great challenge in terms of AP selection. Signal strength strategy (SSS), which selects the AP that provides the highest received signal power for the user equipment (UE), is widely employed in homogeneous networks where the coverage overlap is restricted among APs. In this scenario, load balancing (LB) is optional when the traffic loads are distributed unevenly in geography, and it only applies to the cell-edge UEs which are covered by more than one AP. However, severe traffic imbalance could occur when the SSS method is applied to hybrid networks \cite{ye2013user}, even though the traffic loads are distributed evenly in geography. As a result, the effectiveness of SSS is drastically compromised, while LB becomes an essential and challenging resource management issue in HLWNets.

\begin{table*}[t]
\renewcommand{\arraystretch}{1.3}
\centering
\small
\caption{A summary of related work on resource allocation in HetNet. (UC: User-centric, \checkmark: Supported, \ding{53}: Not supported, N/A: Not applicable)}
\label{Table: Literature Review}
\begin{tabular}{|c|c|c|c|c|c|c|c|c|} 
\hline
\multirow{2}{*}{\textbf{Ref.}} & \multirow{2}{*}{\textbf{Envir.}} & \multirow{2}{*}{\textbf{Method}} & \multicolumn{2}{c|}{\textbf{Function}} & \multicolumn{4}{c|}{\textbf{Performance \& Cost}} \\ 
\cline{4-9}
& & & {\textbf{LB}} & {\textbf{UC}} & \textbf{Optimality} & {\textbf{Complexity}} & {\textbf{Update Interval}\footnote{}} & {\textbf{Feedback}} \\ 
\hline
\cite{li2015cooperative} & \multirow{7}{*}{\vtop{\hbox{\strut Quasi}\hbox{\strut -static} }} & Global optimisation & \checkmark & \ding{53} & Optimal & Very high & \multirow{7}{*}{N/A} & \multirow{6}{*}{High}\\ 
\cline{1-1} \cline{3-7}
\cite{wang2017load} & & Game theory & \checkmark & \ding{53} & \multirow{2}{*}{Near-optim.} & \multirow{2}{*}{High} &  &  \\   
\cline{1-1} \cline{3-5}
\cite{aboagye2021joint} & & College admission model & \checkmark & \ding{53} & &  &  & \\  
\cline{1-1} \cline{3-7}
\cite{wu2017access} & & Fuzzy logic & \checkmark & \ding{53} & Sub-optim. & Low & &  \\  
\cline{1-1} \cline{3-6} \cline{7-7}
\cite{ji2022novel} & & Mixed fuzzy logic and optim. & \checkmark & \ding{53} & \multirow{3}{*}{Near-optim.}  & Medium &  &  \\  
\cline{1-1} \cline{3-5} \cline{7-7}
\cite{alenezi2020reinforcement,ahmad2021reinforcement} &  & Reinforcement learning & \checkmark & \ding{53} & & \multirow{2}{*}{Ultra low} &  &  \\  
\cline{1-1} \cline{3-5} \cline{9-9}
\cite{ji2023adaptive} &  & Deep neural network & \checkmark & \checkmark & &  & & Medium \\ 
\hline 
\cite{wang2015dynamic} & \multirow{7}{*}{Mobile} & Decomposition-based optim. & \checkmark & \ding{53} & Optimal & \multirow{2}{*}{High} & Fixed (500ms) & \multirow{3}{*}{High} \\  
\cline{1-1} \cline{3-6} \cline{8-8}
\cite{li2016mobility} &  & College admission model & \checkmark & \ding{53} & Near-optim. &  & \multirow{2}{*}{Fixed} &  \\  
\cline{1-1} \cline{3-7}
\cite{8943127} &  & Fuzzy logic & \checkmark & \ding{53} & Sub-optim. & Low & & \\    
\cline{1-1} \cline{3-6} \cline{7-9}
\cite{wu2020smart} &  &  Handover skipping & \ding{53} & \checkmark & \multirow{3}{*}{N/A} 
& \multirow{4}{*}{Ultra low} & \multirow{2}{*}{Fixed TTT (160ms)} & \multirow{2}{*}{Medium} \\ 
\cline{1-1} \cline{3-5}
\cite{wu2020novel} &  & ANN-aided handover & \ding{53} & \checkmark & & & &  \\ 
\cline{1-1} \cline{3-5} \cline{8-9} 
\cite{ma2022artificial} &  & ANN-aided handover & \ding{53} & \checkmark & &  & Fixed TTT (2s) & \multirow{2}{*}{Low} \\ 
\cline{1-1} \cline{3-6} \cline{8-8} 
This work &  & Deep neural network & \checkmark & \checkmark & 
Near-optim. & & Adaptive & \\
\hline
\end{tabular}
\end{table*}

Another significant challenge that is faced by HLWNets is mobility management. For a mobile UE, handovers are necessary to transfer it from one AP to another, in order to maintain a decent link quality. There are two basic types of handovers in hybrid networks: i) horizontal handover (HHO), which takes place between two APs of the same type; and ii) vertical handover (VHO), which occurs between different types of APs. Due to the relatively small coverage of LiFi APs, the HHOs within LiFi could be very frequent for fast-moving UEs, compromising their wireless connectivity. Unlike HHOs, VHOs are usually caused by the demands for LB. As the LB solution of one UE is affected by other UEs, frequent VHOs might still occur when the UE is moving slow or even static. This makes mobility management become a tricky issue for HLWNets, particularly in conjunction with the LB problem stated above.

\footnotetext[1]{The update interval is the amount of time between two consecutive implementations of the algorithm. For the handover schemes, the update interval is in accordance with the channel coherence time, which varies with carrier frequency, while a fixed hysteresis margin named the time-to-trigger (TTT) is usually adopted to prevent frequent handovers. Note that the amount of TTT is an order of magnitude larger than the channel coherence time. Hence, the update interval here refers to TTT for the handover schemes.}

\subsection{Related Work}

\subsubsection{Load Balancing in a Quasi-Static Environment} The LB problem is comprised of two sub-problems for the UE: i) selecting a host AP and ii) allocating resource of the host AP. A joint optimisation problem is usually formulated, which establishes a mathematical model between the input (e.g., channel quality, resource availability, quality of service (QoS) requirements, etc.) and the output (i.e., AP selection and resource allocation). So far, a considerable number of studies have been carried out to investigate LB for HLWNets in a quasi-static environment \cite{li2015cooperative,wang2017load,aboagye2021joint,wu2017access,ji2022novel,alenezi2020reinforcement,ahmad2021reinforcement,ji2023adaptive}, and these studies can be classified into two broad categories: i) physical model-based methods and ii) machine learning-based methods. The former category consists of global optimisation, iterative algorithms, and rule-based algorithms. Among them, global optimisation \cite{li2015cooperative} and iterative algorithms \cite{wang2017load,aboagye2021joint} can reach an optimal (or near-optimal) solution at the cost of exceeding computational complexity. To lower the processing power, rule-based algorithms such as fuzzy logic \cite{wu2017access} were reported to make a direct decision on the LB solution, sacrificing the optimality to some extent. In \cite{ji2022novel}, a compound method that mixes fuzzy logic and optimisation was proposed to enhance the algorithm's optimality. Nevertheless, this method is still sub-optimal as well as sensitive to network deployment. 

In contrast to the physical model-based methods, machine learning-based methods can better balance optimality and complexity, at the expense of a nontrivial training process. A few efforts, e.g., \cite{alenezi2020reinforcement,ahmad2021reinforcement}, have been made to explore reinforcement learning for tackling the LB issue. Though these methods can achieve satisfactory optimality with low complexity, they would need to retrain the models when the UE number changes, greatly limiting its practicability. In our previous work \cite{ji2023adaptive}, a deep neural network (DNN)-based learning method was proposed to embrace a unique feature of determining the LB solution for a single target UE. The method, which is named adaptive target-condition neural network (ATCNN), involves an adaptive mechanism that can map any smaller number of UEs to a preset number, without affecting the LB decision for the target UE. As a result, ATCNN can tackle the LB problem for various UE numbers without the need for retraining. In addition, the ATCNN method can achieve near-optimal network performance with an ultra-low inference time in sub-milliseconds \cite{ji2023adaptive}.

\subsubsection{Load Balancing in a Mobile Environment}
The above papers all consider a quasi-static environment, where the LB solution is determined upon invariant channel knowledge. In a mobile environment, however, the handover process is necessary to transfer the UE from one AP to another. To fit time-varying channels, the LB solution should be recalculated and updated regularly, including both the change of the host AP and the reschedule of AP resources. Most of the existing LB methods \cite{li2015cooperative,wang2017load,aboagye2021joint,wu2017access,ji2022novel,alenezi2020reinforcement,ahmad2021reinforcement} are network-centric, i.e., they rely on a central unit to make the LB decision for all UEs at the same time. Consequently, the LB decision needs to be updated at the same pace for all UEs, regardless of their moving status. However, the UEs would require different update frequencies in accordance to their moving status. For fast-moving UEs, a high update frequency is necessary to retain a strong link connectivity. In contrast, slow-moving UEs might experience unnecessary feedback costs as well as frequent handovers when the update frequency is high. While a few relevant studies have considered a mobile environment \cite{wang2015dynamic,li2016mobility,8943127}, the distinctive demands for the update frequency cannot be met due to the intrinsic nature of network-centric LB methods.

A handful of handover schemes have been studied for HLWNets, from the perspective of user-centric rather than network-centric \cite{wu2020smart, wu2020novel,ma2022artificial}. In \cite{wu2020smart}, the authors amended the standard handover scheme of the 3rd generation partnership project (3GPP) to realise handover skipping, which can effectively suppress frequent handovers in HLWNets. Several learning-aided approaches have also been developed on this topic. In \cite{wu2020novel}, an artificial neural network (ANN) model was built to determine the host AP through balancing several factors including channel quality, resource availability, and UE mobility. A similar work was carried out in \cite{ma2022artificial}, except that the output of the learning model was constructed as a binary classification problem, which determines whether the UE should be transferred to a certain target AP or stay with the current host AP. Unlike the noted LB methods, the handover schemes \cite{wu2020smart, wu2020novel,ma2022artificial} make a decision for each individual UE separately, while a fixed time-to-trigger (TTT) is usually adopted to avoid the ping-pong effect. As a result, these schemes can adapt to the various moving status of UEs. However, they fail to render the crucial capability of LB, leading to an inefficient use of network resources. The related works are summarised in Table \ref{Table: Literature Review}, with a comparison of their functionality and performance. So far, to the best knowledge of the authors, no LB approach has yet been developed to address the distinctive demands for update frequency among the mobile UEs.

\subsection{Contribution}
In this paper, a novel user-centric learning approach is developed to tackle the challenging LB issue for HLWNets in a mobile environment. The proposed method is constructed on the basis of our previous work ATCNN \cite{ji2023adaptive}, which embraces a unique property of user-centric LB. In this work, the update interval of ATCNN is modelled to be adjustable for each individual UE, based on its link status and moving status. The objective is to acquire a suitable update interval for the UE, bridging the needs for resource management and mobility management. The main contributions are three-fold:

\begin{itemize}
\item A DNN model is developed to determine the update interval for an individual UE, with four inputs related to the UE: the current host AP, link quality, movement direction and speed. A throughput-degradation criterion is defined for collecting the ground-truth labels of update interval, which allows the UE throughput to reduce by a certain percentage against an ideal case where the update interval is infinitesimal. This criterion reflects a trade-off between the demands for LB and mobility. The proposed DNN model, which is referred to as mobility-supporting neural network (MSNN), enables the update interval to adapt to the UE moving status.

\item A mobility-supporting LB framework is built, which interconnects ATCNN and the proposed MSNN. This framework is termed as mobility-supporting ATCNN (MS-ATCNN). In general, ATCNN and MSNN form an iterative loop: i) ATCNN determines the host AP for an individual UE and forwards this information to MSNN; and ii) MSNN decides an update interval for implementing ATCNN in the next time. It is worth noting that there is no error propagation in the iterative process, since both outputs of ATCNN and MSNN depend on the current status of the UE. The proposed MS-ATCNN enables the UEs to update their LB solutions at different paces in a mobile environment.

\item The learning model performance of MS-ATCNN is investigated, including ablation studies to verify its effectiveness. Also, the wireless network performance of MS-ATCNN is analysed in a mobile environment, in comparison with several baseline methods including: i) ATCNN without MSNN, ii) conventional network-centric LB methods (such as game theory (GT)), and iii) straightforward AP selection methods without the capability of LB (such as SSS). Results show that MSNN can boost the throughput of ATCNN by up to 38\%, against the case without MSNN. In comparison to GT and SSS, MS-ACTNN can improve the throughput by up to 215\% and 310\%, respectively.

\end{itemize}

The remainder of this paper is organised as follows. The system model of HLWNet is introduced in Section \ref{sec:sys_model}. The proposed method is presented in Section \ref{sec:proposed_method}, including the MSNN model, the MS-ATCNN framework, the criterion for sample data collection, and the process of training and test. Ablation studies of MS-ATCNN are carried out in Section \ref{sec:ablation}. Simulation results are presented in Section \ref{sec:simulation}. Finally, conclusions are drawn in Section \ref{sec:conclusion}.

\section{System Model} \label{sec:sys_model}
The system model of HLWNets in a mobile environment is introduced in this section, including the network model, the channel model, and the mobility model.

\subsection{Network Model}      
\begin{figure}[t]
\centering
\includegraphics[height=3.6in,width=3in] {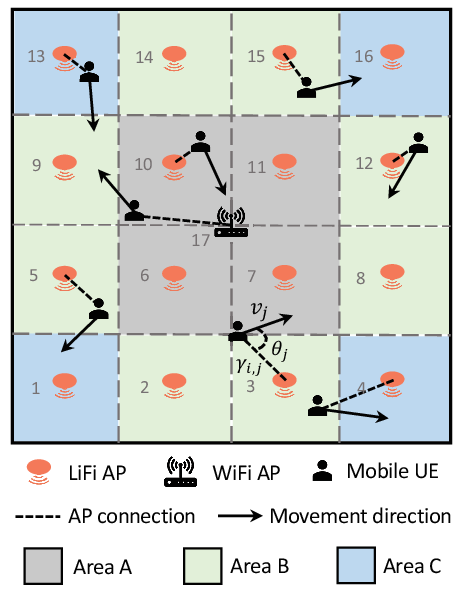} 
\caption{Schematic diagram of an indoor HLWNet (top view).}  
\label{Fig: SystemModel}
\end{figure}

The network topology of the HLWNet considered in this paper is presented in Fig. \ref{Fig: SystemModel}, which consists of 16 LiFi APs and 1 WiFi AP. The LiFi APs are arranged in a grid on the ceiling, while the WiFi AP is located at the centre of the room on the floor. According to the geographic symmetry, the LiFi APs can be divided into 3 areas, which are marked with different colors in Fig \ref{Fig: SystemModel}. Details about this division will be explained in Section \ref{sec: MSNN}. The UEs move around in the room with random initial positions, which are assumed to be uniformly distributed. Let $N_a$ and $N_u$ denote the number of APs and the number of UEs; let $\mathbb{S}=\{1,2,...,N_a\}$ and $\mathbb{U}=\{1,2,...,N_u\}$ denote the set of APs and the set of UEs; let $i$ and $j$ denote the index of APs and the index of UEs, where $i\in \mathbb{S}$ and $j\in \mathbb{U}$, respectively. Each UE has a certain data rate requirement, which is denoted by $R_j$. The data rate requirements across the UEs are assumed to follow a Gamma distribution with unity shape parameter. Let a binary variable $\chi_{i,j}$ indicate the connection status of the link between AP $i$ and UE $j$, then $\chi_{i,j}=1$ means that UE $j$ is connected to AP $i$, and otherwise $\chi_{i,j}=0$. Each UE is served by one and only one AP, which each AP can serve multiple UEs via time-division multiple access. The proportion of time that AP $i$ allocates to UE $j$ is denoted by $\rho_{i,j}$, which is a continuous variable in the range between 0 and 1. 

\subsection{Channel Model}
The LiFi channel is comprised of two components: line-of-sight (LoS) and non-line-of-sight (NLoS) paths, which are denoted by $H_{{\rm{LoS}}}^{i,j}$ \cite[eq. (1)]{ji2022novel} and $H_{{\rm{NLoS}}}^{i,j}$ \cite[eq. (3)]{ji2022novel}, respectively. The overall LiFi channel is expressed as $H_{\text{LiFi}}^{i,j} = H_{{\rm{LoS}}}^{i,j} + H_{{\rm{NLoS}}}^{i,j}$. Let $\gamma_{\rm{LiFi}}^{i,j}$ denote the signal-to-interference-plus-noise ratio (SINR) of the LiFi link, which can be formulated as follows: 
\begin{equation} 
\gamma_{\rm{LiFi}}^{i,j} =  \frac{(R_{\rm{pd}}H_{\rm{LiFi}}^{i,j}P_{\rm{mod}})^2}{N_{\rm{LiFi}}B_{\rm{LiFi}}+\sum\limits_{{i'\in \mathbb{S}},{i'\neq{i}}}(R_{\rm{pd}}H_{\rm{LiFi}}^{i',j}P_{\rm{mod}})^2},   
\end{equation}  
where $N_{\rm{LiFi}}$ is the noise power spectral density (PSD) of LiFi; $B_{\rm{LiFi}}$ denotes the channel bandwidth of LiFi; $R_{\rm{pd}}$ is the detector responsivity; and $P_{\rm{mod}}$ stands for the modulated optical power.

The WiFi channel is denoted as $H_{\rm{WiFi}}^{i,j}$, which follows the channel model in \cite[eq. (9)]{wu2017access}. Since a single WiFi AP is involved in this paper, co-channel interference is neglected. Let $\gamma_{\rm{WiFi}}^{i,j}$ denote the signal-to-noise ratio (SNR) of the WiFi link between AP $i$ and UE $j$, and it can be given by:
\begin{equation} 
\gamma_{\rm{WiFi}}^{i,j} =  \frac{\left|H_{\rm{WiFi}}^{i,j}\right|^{2}P_{\rm{WiFi}}}{N_{\rm{WiFi}}B_{\rm{WiFi}}},   
\end{equation}
where $P_{\rm{WiFi}}$ is the transmit power of the WiFi AP; $N_{\rm{WiFi}}$ is the noise PSD of WiFi; and $B_{\rm{WiFi}}$ denotes the channel bandwidth of WiFi.

The link capacity between AP $i$ and UE $j$ is denoted by $C_{i,j}$, which can be expressed as follows \cite[eq. (1)]{ji2023adaptive}:
\begin{equation}
C_{i,j} =
\begin{cases}
  \dfrac{{B_{\rm{LiFi}}}}{2}{\log_2}\left(1 + {\dfrac{e}{2\pi}} \gamma_{\rm{LiFi}}^{i,j}\right),  & \forall i \in
  {\mathbb{S}}_{\rm{LiFi}}\\ 
  {B_{\rm{WiFi}}}{\log_2}(1 + \gamma_{\rm{WiFi}}^{i,j}) , & \forall i \in {\mathbb{S}}_{\rm{WiFi}}  \\ 
\end{cases},
\end{equation}
where $e$ is the Euler’s number; ${\mathbb{S}}_{\rm{LiFi}}$ is the set of LiFi APs; ${\mathbb{S}}_{\rm{WiFi}}$ is the set of WiFi APs, despite only one WiFi AP is involved in this paper.

\subsection{Mobility Model}\label{Section:Mobility Model}
Commonly used synthetic mobility models for wireless networks include random walk, random waypoint (RWP), and Gauss-Markov mobility models \cite{camp2002survey}. The random walk and RWP are memoryless mobility models, where the current value of velocity is independent of its history values. On the contrary, the Gauss-Markov model generates time-correlated velocities, avoiding unrealistic sudden stops and turns in the memoryless models. Among those models, the RWP mobility model is mostly adopted in the research on wireless communications, e.g., \cite{wang2015dynamic,li2016mobility,8943127,wu2020smart,wu2020novel,ma2022artificial,wu2019mobility-aware,wu2020parallel,ahmad2022sequential,chatterjee2024mobility}.

With the RWP model, UE $j$ moves in a straight line from one point $P_j^{(l)}$ to the next point $P_j^{(l+1)}$, where the points are randomly located in the room area following a uniform distribution. During this excursion, UE $j$ adopts a constant speed $v_j^{(l)}$, which is uniformly distributed in range $\left(0, v_{\rm{max}}\right]$. Let $\theta_{j}^{(l)}$ represent the angle between the UE's movement direction and the line that connects the UE to its host AP, as demonstrated in Fig. \ref{Fig: SystemModel}.
The UE movement leads to a time-varying channel and hence a time-varying link capacity. Let $C_{i,j}^{(t)}$ denote the link capacity between AP $i$ and UE $j$ at time instance $t$. Similar notations apply to $\rho_{i,j}^{(t)}$ and $\chi_{i,j}^{(t)}$. The instantaneous throughput achieved by UE $j$ at time instance $t$ is denoted by $\Gamma_{j}^{(t)}$, which can be calculated as follows:
\begin{equation}\label{Eq:UE_Throughput}
\Gamma_{j}^{(t)} = \sum\limits_{i \in \mathbb{S}}{\rho_{i,j}^{(t)} }{\chi_{i,j}^{(t)}}{C_{i,j}^{(t)}}.
\end{equation}

\section{Proposed Method for Resource and Mobility Management} \label{sec:proposed_method}

\begin{figure*}[h]
\centering
\includegraphics[width=7.2in,height=2.4in]{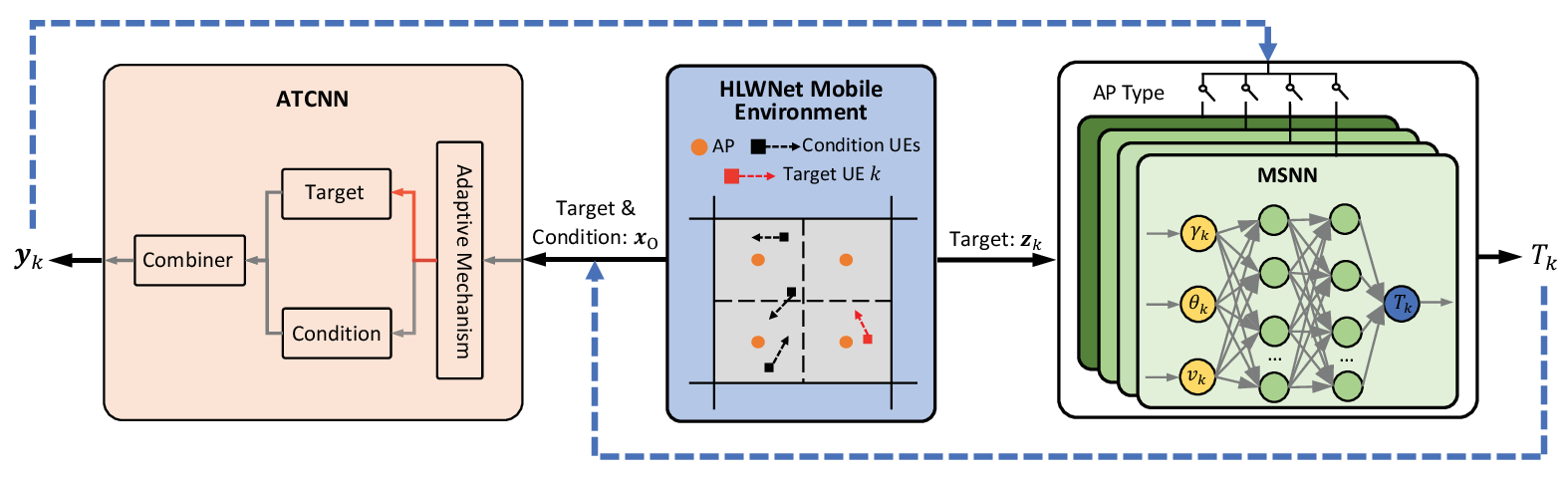}
\caption{Schematic diagram of the proposed MS-ATCNN (solid lines: information flow; dashed lines: control flow).} 
\label{Fig: MS-ATCNN}
\end{figure*}

In this section, the proposed MSNN model is elaborated, followed by an introduction to the framework of MS-ATCNN, which interconnects MSNN and ATCNN to jointly tackle the resource and mobility management. Then the throughput-degradation criterion is introduced for collecting the sample dataset of update intervals. Finally, we present the process of training, validation, and test.

\subsection{Mobility-Support Neural Network (MSNN)} \label{sec: MSNN}
The MSNN model aims to yield an update interval that suits the mobile UE, based on its status including the SNR, movement direction, and speed. These variables are denoted by $\gamma_{k}$, $\theta_{k}$, and $v_{k}$, respectively, where $k$ denotes the target UE. The output is a single variable $T_{k}$, which stands for the update interval that is estimated for the target UE. The MSNN model may differ from one AP to another, due to the different locations and coverage areas. Specifically, the WiFi AP covers a much larger area than the LiFi AP, resulting in distinctive demands for the update interval. Though the LiFi APs have the same coverage area, their locations are different, which can also lead to variations in the required update interval. For this reason, the APs are classified into several types, and each type employs a dedicated MSNN model. 

Regarding the HLWNet in Fig. \ref{Fig: SystemModel}, there are three kinds of LiFi APs depending on their locations (which are divided by Areas A, B, and C), in addition to a single WiFi AP. Thus, the APs are classified into four types: AP Type I contains 4 LiFi APs $\langle6, 7, 10, 11\rangle$ located in Area A; AP Type II contains 8 LiFi APs $\langle2, 3, 5, 8, 9, 12, 14, 15\rangle$ located in Area B; AP Type III contains the remaining 4 LiFi APs $\langle1, 4, 13, 16\rangle$ located in Area C; and AP Type IV is the WiFi AP $\langle17\rangle$. It is worth noting that the AP classification is subject to the symmetry of network topology. As for an asymmetrical topology, each AP can be an individual type. In this paper, the MSNN model is trained for just one AP in each type, since the trained MSNN model can be directly applied to the other APs in the same type. In practice, each AP can implement its own MSNN model, for the purpose of generality.

A DNN structure is designed for MSNN to map the input $\boldsymbol{z}_{k}=[\gamma_{k}, \theta_k, v_k]$ to the output $T_{k}$, as shown in Fig. \ref{Fig: MS-ATCNN}. Specifically, $\boldsymbol{z}_{k}$ is fed to two fully connected hidden layers. Let $N_{L1}$ denote the number of neurons in the first layer and $N_{L2}$ for the second layer. In the first layer, the weight matrix and the bias vector are denoted by $\mathbf{W}_1 \in \mathbb{R}^{3 \times N_{L1}}$ and $\boldsymbol{b}_1 \in \mathbb{R}^{N_{L1}}$. Similarly, the weight matrix and the bias vector are $\mathbf{W}_2 \in \mathbb{R}^{N_{L1} \times N_{L2}}$ and $\boldsymbol{b}_2 \in \mathbb{R}^{N_{L2}}$ for the second layer. The activation function adopted for the hidden layers is a rectified linear unit (ReLU), which is defined as ${f_{\rm{R}}}(\boldsymbol{x}) = {\rm{max}}({\rm 0}, \boldsymbol{x})$. In the output layer, the weight matrix and the bias vector are denoted by $\mathbf{W}_3 \in \mathbb{R}^{{N_{L2}} \times 1}$ and $\boldsymbol{b}_3 \in \mathbb{R}^{1}$, respectively. Afterward, the sigmoid activation function is used, which is defined as ${f_{\rm{S}}}(\boldsymbol{x}) = 1 / (1+e^{\boldsymbol{-x}})$. Finally, the MSNN model outputs the estimated update interval $T_k$ as follows:
\begin{equation}
T_k = {f_{\rm{S}}}\Bigl( {\mathbf{W_3}} {{f_{\rm{R}}}\bigl({{\mathbf{W_2}} {{f_{{\rm{R}}}}\left( {{\mathbf{W_1}}{\boldsymbol{z}_k} + {\boldsymbol{b_1}}} \right)} + \boldsymbol{b_2}} \bigl)} + \boldsymbol{b_3} \Bigl).
\end{equation}

\subsection{Framework of MS-ATCNN} \label{sec:MS-ATCNN}

Fig. \ref{Fig: MS-ATCNN} outlines the framework of MS-ATCNN, which consists of three key components: i) the MSNN model, ii) the ATCNN model, and iii) the mobile environment of an HLWNet. For a certain target UE, MS-ATCNN runs the resource and mobility management in a loop, which involves four steps.
\begin{itemize}
\item In the first step, information about the UEs is extracted from the mobile environment and fed to the ATCNN model. For UE $j$, the information is denoted by $\boldsymbol{x}_{j} = [\gamma_{1,j}, \gamma_{2,j}, ..., \gamma_{N_a,j}, R_j]$, which is comprised of the SNR values and the required data rate. The overall information is denoted by $\boldsymbol{x}_{\rm{O}} = [{\boldsymbol{x}_1}, {\boldsymbol{x}_2}, ..., {\boldsymbol{x}_k}, ..., {\boldsymbol{x}_{N_u}}]$, which contains the relevant information on both the target UE $k$ and the other UEs (which are referred to as condition UEs) in the mobile environment.

\item In the second step, the ATCNN determines the host AP for the target UE, upon the information it receives. As shown in Fig. \ref{Fig: MS-ATCNN}, the ATCNN model consists of four function blocks: i) an adaptive mechanism to map any smaller number of UEs to a preset number; ii) a target neural network to extract the features of the target UE; iii) a condition neural network to extract the features of the condition UEs; and iv) a combiner neural network to make an LB solution for the target UE. See details about the ATCNN model in our previous work \cite{ji2023adaptive}. The output of ATCNN is denoted by $\boldsymbol{y}_k$, which indicates the host AP for the target UE. If $\boldsymbol{y}_k$ differs from the previous result, the UE will be handed over to a new host AP; otherwise, the UE will stay with the current host AP until the next implementation of ATCNN.  

\item In the third step, the MSNN model is run to decide an update interval for the target UE. Upon the latest result from ATCNN, the corresponding AP implements its own MSNN. As mentioned, there are four types of APs in the considered HLWNet. Therefore, Fig. \ref{Fig: MS-ATCNN} presents a switch among four MSNN models, for illustration purposes. The MSNN model also acquires information from the mobile environment. Details about the MSNN model have been introduced in the previous subsection.

\item In the last step, the output of the MSNN model (i.e., the update interval) is used to control when the ATCNN model is implemented for the next time. This is realised through controlling the information flow that is fed to ATCNN, i.e., the first step.

\end{itemize}

\begin{figure}[t]
\centering
\includegraphics[height=4in, width=2.6in]{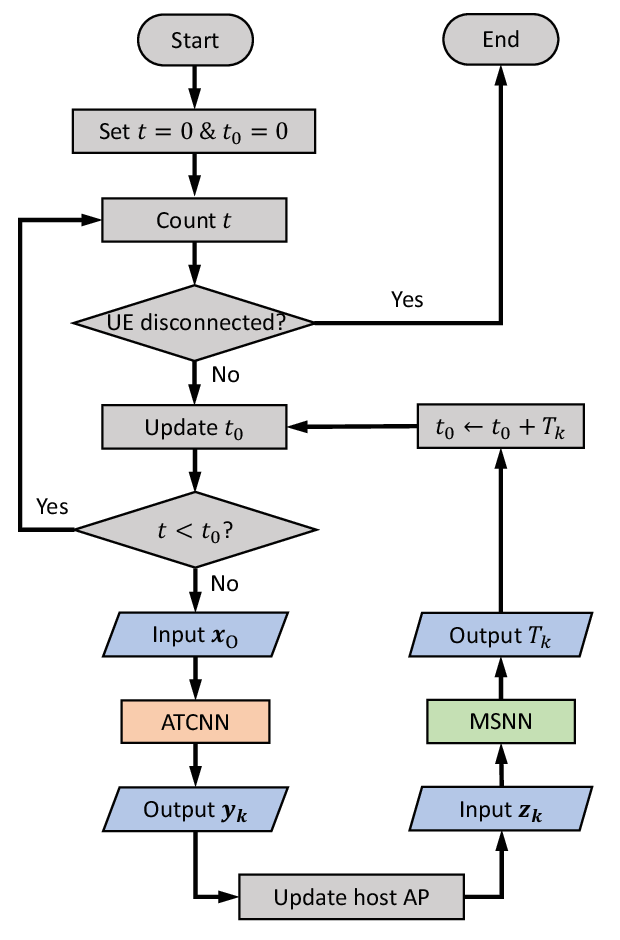}
\caption{Flowchart of the proposed MS-ATCNN.}
\label{Fig: Flowchart}
\end{figure}

Fig. \ref{Fig: Flowchart} presents the flowchart of the proposed MS-ATCNN scheme, and its pseudocode is given in Alg. 1. In general, the ATCNN model copes with the resource management, while the MSNN model handles the mobility management. The two models interact with each other, forming the concept of MS-ATCNN. Since the MS-ATCNN scheme operates independently for each individual UE, adaptive update intervals are feasible among the UEs, allowing a more flexible way of tackling the resource and mobility management than the conventional network-centric LB methods.

\subsection{Dataset Collection}\label{sec:Dataset_Collection}
To train the proposed MSNN model, it is necessary to collect sample data of the update interval under a certain criterion. Since throughput is the key metric to measure the effectiveness of LB, we introduce a throughput-degradation criterion to obtain the ground-truth labels of update interval, which allows the throughput of the target UE to drop by a certain percentage against the ideal case with an infinitesimal update interval. For practical implementations, here the ideal ATCNN is defined as running the ATCNN model at a fixed update interval of 10ms, considering the channel coherence time in HLWNets.

\begin{algorithm}[t]
\SetAlgoLined\label{Algorithm}
\textbf{Initialisation:} $t \gets 0$, $t_0 \gets 0$, $\delta \gets $1ms\;
\While{\rm{UE} $k$ is connected to the network}{
\If{$t = t_0$}{
Extract $\boldsymbol{x}_{\rm{O}}^{(t)}$ and $\boldsymbol{z}_{k}^{(t)}$ from the network\;
$\boldsymbol{y}_k^{(t)} \gets$ Input $\boldsymbol{x}_{\rm{O}}^{(t)}$ to ATCNN\; 
\If{$\boldsymbol{y}_k^{(t)}$ \rm{differs from previous}}{Handover UE $k$ to the new host AP $\boldsymbol{y}_k^{(t)}$\;$\Phi_{k} \gets$ Determine the AP type of $\boldsymbol{y}_k^{(t)}$\;}
$T_k^{(t)} \gets$ Input $\boldsymbol{z}_{k}^{(t)}$ to MSNN with type $\Phi_{k}$\;  
$t_0 \gets t_0 + T_k^{(t)}$\;}
$t \gets t+\delta$\; 
}
\caption{Algorithm of MS-ATCNN}
\end{algorithm}

Let $\boldsymbol{y}_k^{(t)}$ denote the output of ATCNN at time $t$. Let $t_{0}$ denote the reference point of time. Assume that the ATCNN model is implemented at $t_{0}$, and the output $\boldsymbol{y}_k^{(t_{0})}$ is applied to the target UE until $t_{1}$. Based on (\ref{Eq:UE_Throughput}), the average throughput during this period can be calculated by:
\begin{equation}
\Gamma_{k}^{t_{0}\rightarrow t_{1}}\left(\boldsymbol{y}_k^{(t_{0})}\right) = \frac{1}{t_{1}-t_{0}} \int_{t_{0}}^{t_{1}} \left(\sum\limits_{i \in \mathbb{S}}{\rho_{i,j}^{(t)} }{\chi_{i,j}^{(t_{0})}}{C_{i,j}^{(t)}} \right)dt.
\end{equation}

Let $\dot{T}_k$ denote the desired update interval. The average throughput resulted by MS-ATCNN can be expressed as $\Gamma_{k}^{t_{0}\rightarrow \dot{T}_k}\left(\boldsymbol{y}_k^{(t_{0})}\right)$. Meanwhile, the ideal ATCNN updates its output every 10ms. Thus, the average throughput achieved by the ideal ATCNN can be computed by:
\begin{equation}
\Tilde{\Gamma}_{k}^{t_{0}\rightarrow \dot{T}_k}= \dfrac{1}{M} \sum\limits_{m=1}^{M} \Gamma_{k}^{t\rightarrow t+\Delta t}\left(\boldsymbol{y}_k^{(t)}\right),
\end{equation}
where $\Delta t$ is 10ms for the defined ideal ATCNN and $t=t_{0}+(m-1)\Delta t$, while $M$ is the integer division between $\dot{T}_k$ and $\Delta t$. The throughput-degradation percentage, which is denoted by $\Delta\Gamma$, is limited within 5\% in this paper, so that the proposed MS-ATCNN can achieve a satisfactory throughput performance while allowing a flexible update interval to some extent. Accordingly, the desired update interval $\dot{T}_k$ can be obtained by solving the following problem:
\begin{equation} \label{MSNN_formulation} 
	\begin{split}
		\max \quad & \left\{\dot{T}_k \bigg| \Gamma_{k}^{t_{0}\rightarrow \dot{T}_k}\left(\boldsymbol{y}_k^{(t_{0})}\right) \geq \left(1-\Delta\Gamma \right) \times  \Tilde{\Gamma}_{k}^{t_{0}\rightarrow \dot{T}_k} \right\} \\
		\text{s.t.}\quad & \dot{T}_k \in \left[0.01s, 2s\right]; \\
  & \Delta\Gamma \leq 5\%.
	\end{split}
\end{equation}

In practical applications, the dataset collection can be conducted through the following steps. First, the input data can be gathered via standard communication protocols or built-in components in mobile devices. Specifically, the information on the UE's speed and orientation can be measured by the accelerometer and gyroscope, while the SINR data is readily accessed in the process of channel estimation. Second, the input data is forwarded to a network server for creating a digital twin environment, which allows network analysis without affecting the real-world network. Third, 
the server implements ATCNN every 10ms and monitors the network throughput through a network simulator such as Wireshark. In the last step, the ground-truth label of update interval is obtained following the throughput-degradation criteria.

\subsection{Training, Validation and Test}

\subsubsection{Training and Validation} Following the above criterion, 2000 samples are collected to train the MSNN model of each AP type. The dataset is pre-processed with a linear normalisation, before it is fed into the MSNN model with a ratio of 80:20 between training data and validation data. The loss function is mean square error (MSE), which is commonly used for regression tasks in machine learning. The MSE loss function can be expressed as follows:  
\begin{equation}\label{Equation: MSE Loss} 
L_{\mathrm{MSE}}\left( \alpha \right)= \frac{1}{N}{\sum_{n=1}^{N}}{\left(T_{k,n} - \dot{T}_{k,n}\right)^2},  
\end{equation}
where $\alpha$ denotes the set of all the weight matrices and bias vectors in the MSNN model; $N$ is the number of samples; $T_{k,n}$ stands for the estimated update interval for UE $k$ in the $i$-th sample; and $\dot{T}_{k,n}$ denotes the corresponding ground-truth label. The adaptive moment estimation (Adam) \cite{kingma2014adam} is employed to train the MSNN model by iterating $\alpha$ through $ \alpha \gets \alpha -\eta \nabla L_{\mathrm{MSE}}\left ( \alpha  \right )$, where $\eta$ is the learning rate, and $\nabla L_{\mathrm{MSE}}(\alpha)$ stands for the gradient of the loss function with respect to $\alpha$. See details about the setup of the training parameters in Section \ref{sec:simulation}.

\begin{figure}[t] 
\centering
\subfigure[AP Type I.]{\includegraphics[width=0.21\textwidth]{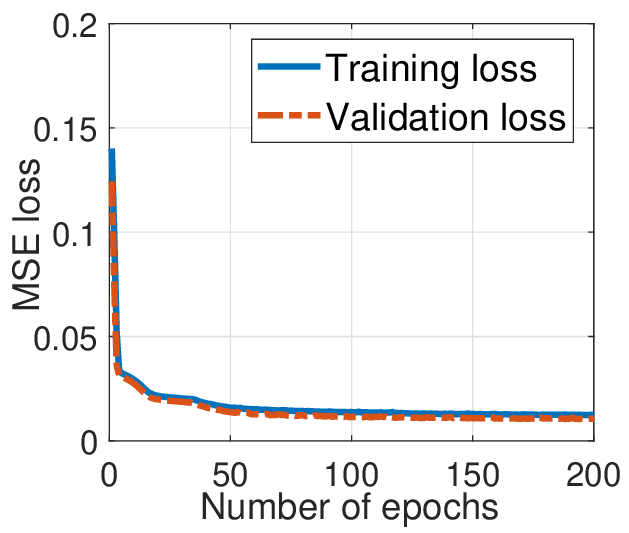}
\label{fig:Loss_TypeI}}
\subfigure[AP Type II.]{\includegraphics[width=0.215\textwidth]{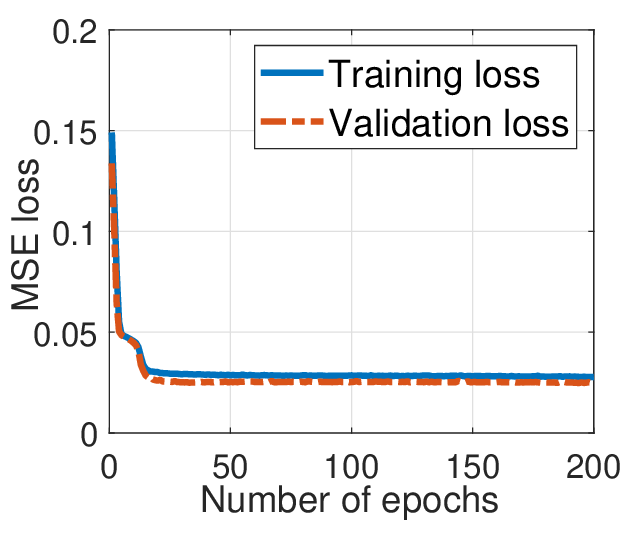}
\label{fig:Loss_TypeII}}
\subfigure[AP Type III.]{\includegraphics[width=0.217\textwidth]{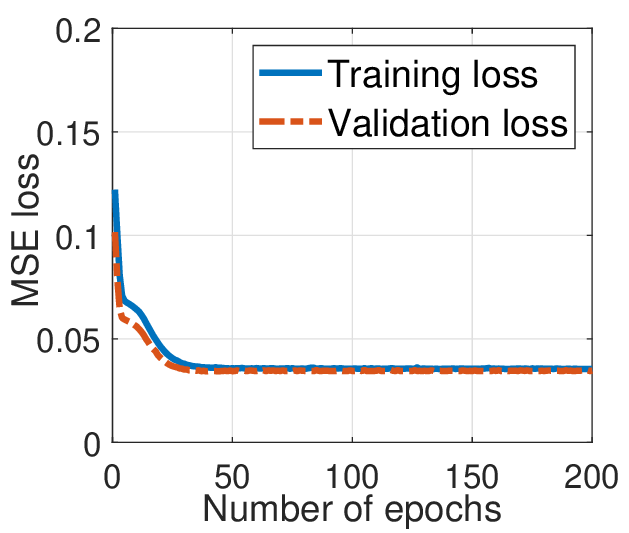}
\label{fig:Loss_TypeIII}}
\subfigure[AP Type IV.]{\includegraphics[width=0.21\textwidth]{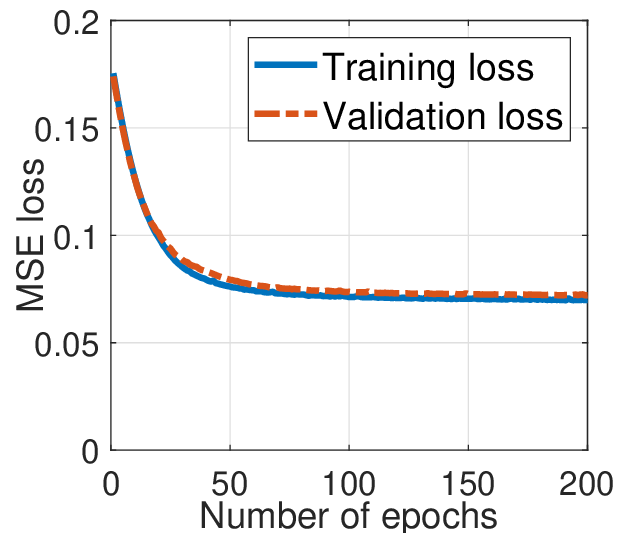}
\label{fig:Loss_TypeIV}}
\caption{Training loss and validation loss of the MSNN model.}
\label{Fig: MSNN Loss}
\end{figure}

Fig. \ref{Fig: MSNN Loss} presents the training loss and validation loss of the MSNN model for different AP types. In all cases, the validation loss matches the training loss closely, indicating that the MSNN model is adequately trained with no evident over-fitting or under-fitting phenomenons. It is worth noting that the steady-state loss of AP Type IV (i.e., the WiFi AP) in Fig. \ref{fig:Loss_TypeIV} is noticeably higher than that of the other AP types (i.e., the LiFi APs) in Fig. \ref{fig:Loss_TypeI}-\ref{fig:Loss_TypeIII}. This is mainly because the WiFi AP has a larger coverage area than the LiFi AP, leading to a less accurate training outcome. Apart from that, the convergence rates are slightly different among the AP types. On average, it requires around 50 epochs to reach a steady state.

\subsubsection{Test} To verify the training performance of the MSNN model, we analyse the cumulative distribution function (CDF) of the throughput gap between MS-ATCNN and the ideal ATCNN, which is defined in Section \ref{sec:Dataset_Collection}. The results are collected through 200 simulations with each lasting for 10s. Note that the ground-truth update intervals might be very long for slow-moving UEs, even exceeding the duration of simulations. For practical implementations, the ground-truth update intervals are capped at 2s. 

\begin{figure}[t]
\centering
\includegraphics[height=2.8in,width=3.2in]{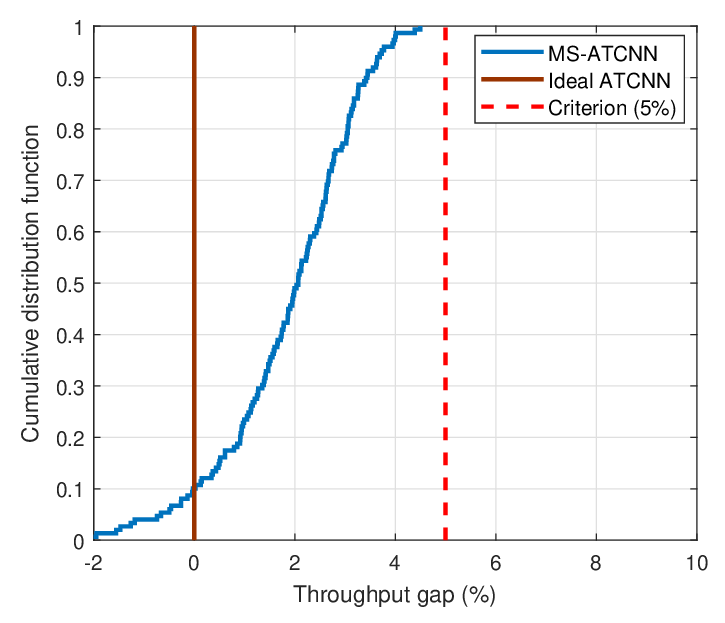}
\caption{Throughput gap between MS-ATCNN and ideal ATCNN.}
\label{Fig: GapCDF}
\end{figure}

As shown in \mbox{Fig. \ref{Fig: GapCDF}}, the throughput gap of MS-ATCNN is well below the preset throughput-degradation criterion of 5\%. In fact, the average throughput gap is just around 2\%. This is because capping the ground-truth update interval alleviates the throughput degradation. It is also observed that the throughput gap of MS-ATCNN has a 10\% probability of being negative. In other words, MS-ATCNN occasionally achieves a higher throughput than the ideal ATCNN. The reason is that the objective of ATCNN is to maximise the overall network throughput with a proportional fairness, rather than the throughput of the target UE. For certain scenarios, it is possible that MS-ATCNN with a longer update interval achieves a higher throughput than the case with a shorter update interval. For instance, when the target UE is moving towards it current host AP, the ATCNN model might transfer the UE to another AP that is further away, from the angle of balancing the overall network data traffic. With a longer update interval, the UE can stay with its current host AP and achieve a higher throughput, at the cost of compromising the throughput performance of other UEs.

\begin{figure}[t]
\centering
\includegraphics[height=2.8in,width=3.2in]{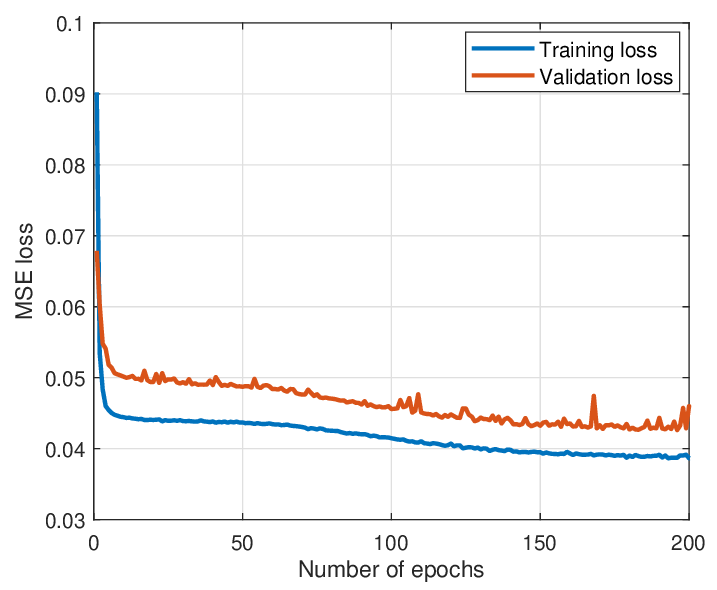}
\caption{Training loss and validation loss of MSNN without AP classification.}
\label{Fig: Ablation_MixedLoss}
\end{figure}

\section{Ablation Study} \label{sec:ablation}
In this section, two ablation studies are implemented to verify the effectiveness of the proposed MSNN model. The first study is examining the necessity of using dedicated MSNN models for different AP types, while the second one is analysing the impact of each input variable on the MSNN model. The simulation setup is the same as in Section \ref{sec:simulation}.

\subsection{Removal of AP Classification}
As mentioned in Section \ref{sec: MSNN}, the APs of the considered HLWNet are classified into four types, and each AP type is associated with a dedicated MSNN model. To verify the necessity of this implementation, we analyse the training performance of the MSNN model without AP classification, where the sample datasets of all AP types are mixed for training the model. Fig. \ref{Fig: Ablation_MixedLoss} shows the training and validation losses in this case. As shown, the validation loss is noticeably higher than the training loss, indicating that the MSNN model is over-fitted without AP classification. By comparing \mbox{Fig. \ref{Fig: Ablation_MixedLoss}} with \mbox{Fig. \ref{Fig: MSNN Loss}}, it can be concluded that AP classification can effectively eliminate the over-fitting issue in the training process of the MSNN model. The reason has been explained in Section \ref{sec: MSNN}.

\begin{figure}[t]
\centering
\subfigure[AP Type I.]{\includegraphics[width=0.225\textwidth]{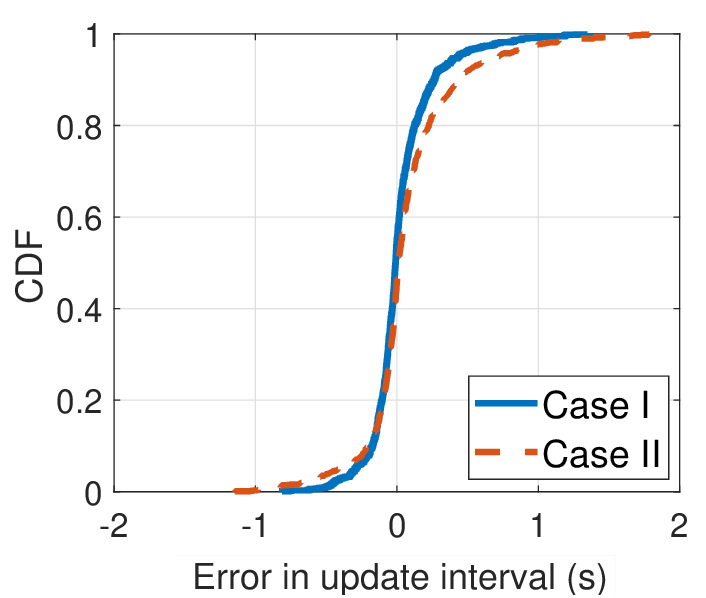}
\label{fig:UTI_TypeI}}
\subfigure[AP Type II.]{\includegraphics[width=0.222\textwidth]{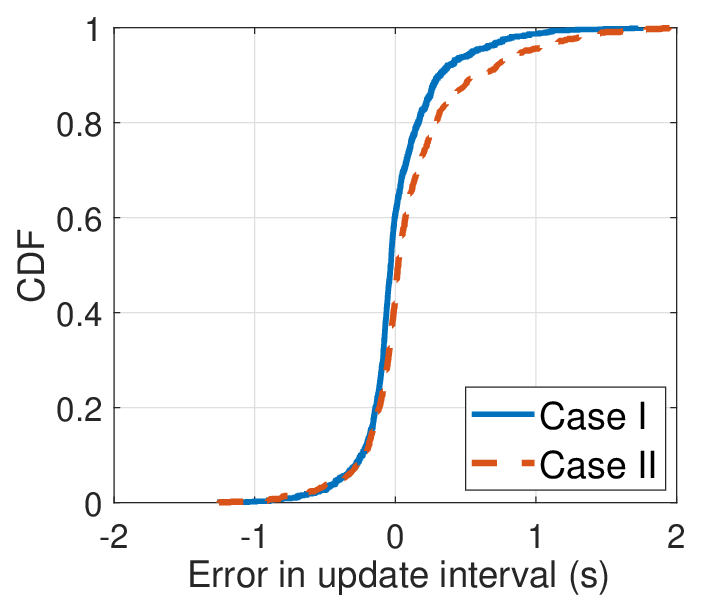}
\label{fig:UTI_TypeII}}
\subfigure[AP Type III.]{\includegraphics[width=0.22\textwidth]{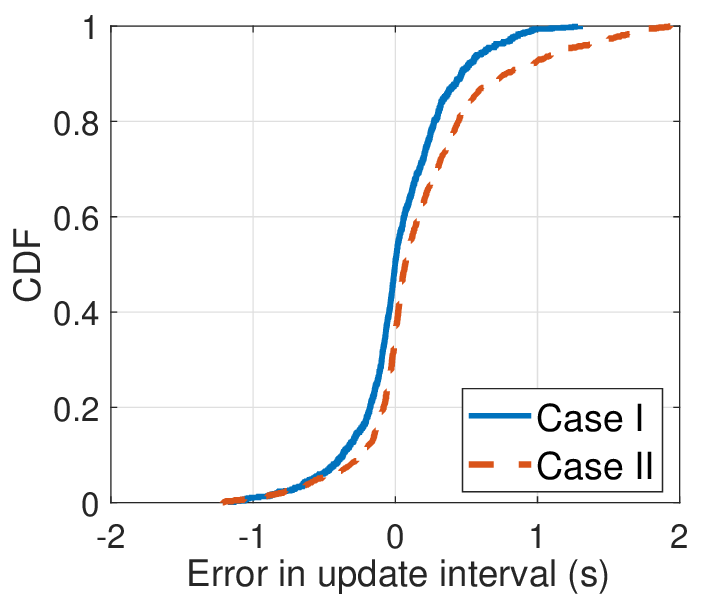}
\label{fig:UTI_TypeIII}}
\subfigure[AP Type IV.]{\includegraphics[width=0.22\textwidth]{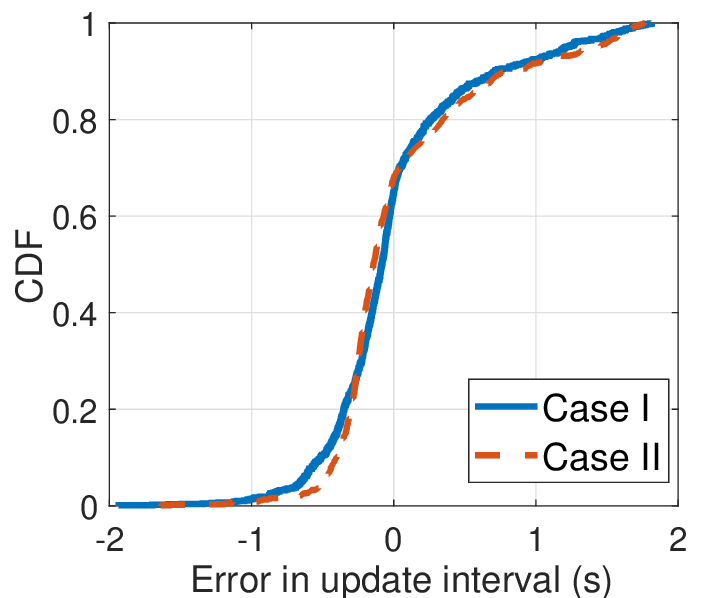}
\label{fig:UTI_TypeIV}}
\caption{Accuracy of the MSNN models with AP classification (Case I) and without AP classification (Case II).}
\label{Fig: update intervalerrors}
\end{figure}

Apart from the training loss, accuracy is also a key metric to measure the performance of a learning model. Here the accuracy is evaluated by the absolute error, which equals the estimated update interval minus the ground-truth label. \mbox{Fig. \ref{Fig: update intervalerrors}} presents the CDF of those errors for the MSNN models with and without AP classification. In general, AP classification makes the MSNN model more accurate for the LiFi AP types (i.e., AP Type I, II and III). As shown, the MSNN models that are dedicated to those types are noticeably closer to zero than the case without AP classification. As for the WiFi AP (i.e., AP Type IV), the two cases exhibit similar levels of accuracy. This is because the WiFi AP covers a larger area than the LiFi AP, leading to a less accurate estimation of the update interval. Consequently, the samples of the WiFi AP dominate the accuracy of the MSNN model when it is trained with the mixed samples of different APs. 

\begin{figure*}[h]
\centering
\subfigure[Removal of SNR.]{\includegraphics[width=0.32\textwidth]{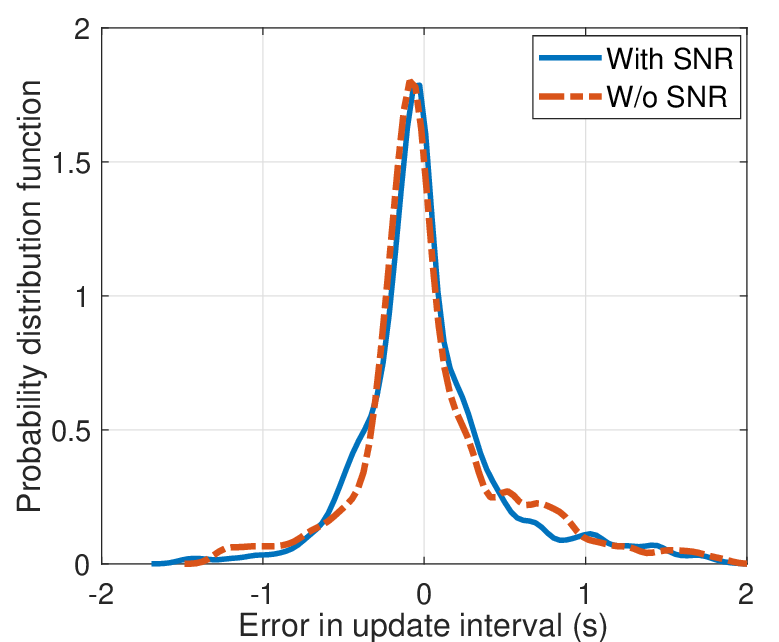}
\label{fig7:ablation_a}}
\subfigure[Removal of movement direction.]{\includegraphics[width=0.32\textwidth]{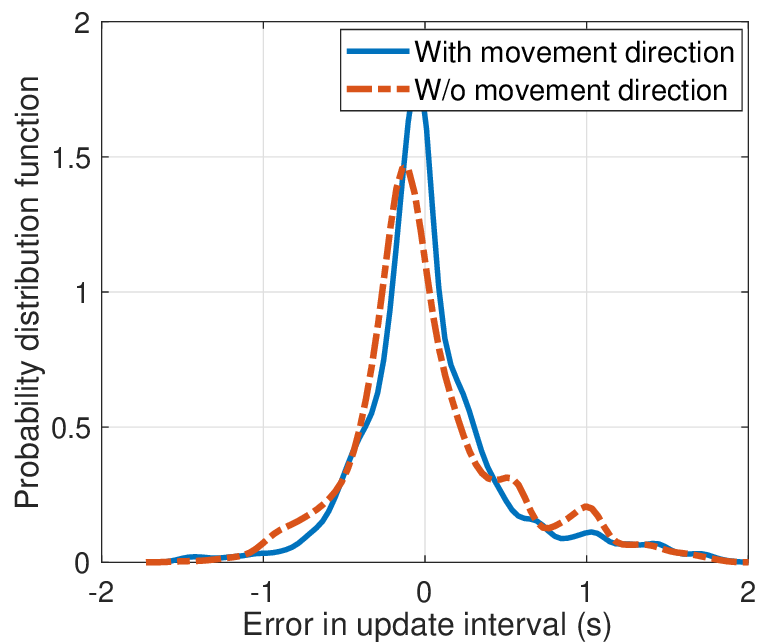}
\label{fig7:ablation_b}}
\subfigure[Removal of speed.]{\includegraphics[width=0.32\textwidth]{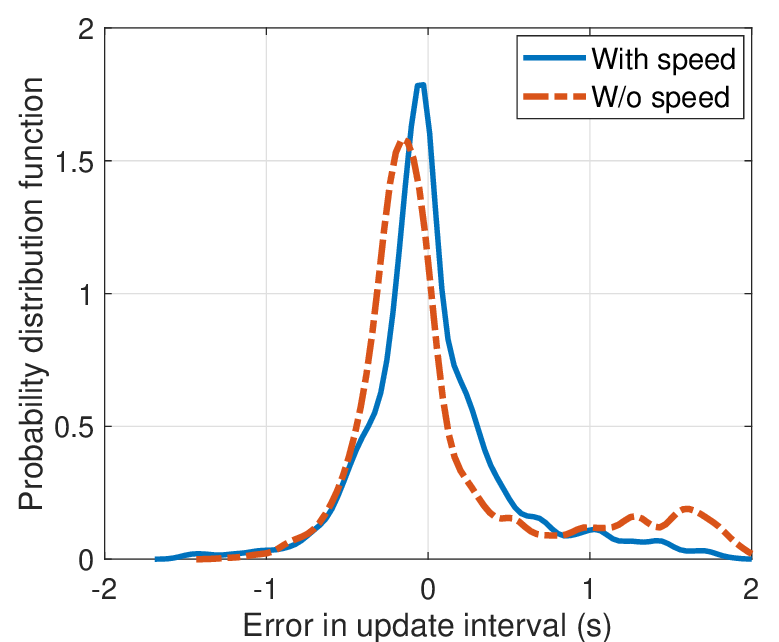}
\label{fig7:ablation_c}}
\caption{Accuracy of the MSNN models with and without individual input variables.}
\label{Fig: Ablation_InputParameters}
\end{figure*}

\subsection{Removal of Individual Input Variables}

Now analyse the accuracy of the MSNN models with and without individual input variables. Fig. \ref{fig7:ablation_a} shows that when SNR is removed from the input, the error variance of MSNN increases from 0.200 to 0.226. Given a confidence level 80\%, the proposed MSNN model obtains a confidence interval (CI) of errors within $[-406\text{ms}, 576\text{ms}]$, which is 11.5\% smaller than the range $[-412\text{ms}, 697\text{ms}]$ offered by the case without SNR. These results signify that involving SNR as an input can help MSNN achieve a higher accuracy. Similarly, when the movement direction is removed from the input, the error variance of the MSNN model increases to 0.236, while the corresponding CI is enlarged to $[-479\text{ms}, 730\text{ms}]$, as shown Fig. \ref{fig7:ablation_b}. A similar but more pronounced trend can be found in the case of removing the variable speed, where the error variance soars to 0.381, as shown in Fig. \ref{fig7:ablation_c}. Meanwhile, the corresponding CI becomes $[-417\text{ms}, 1252\text{ms}]$, which is 70\% larger than the proposed MSNN model. In summary, the three input variables (especially the UE's speed) are effective in delivering an accurate estimation of the update interval.

\section{Simulation Results} \label{sec:simulation}
In this section, Monte Carlo simulations are carried out to evaluate the performance of the proposed MS-ATCNN. Three baseline methods are taken into account: i) ATCNN with different update schemes other than the MSNN model; ii) GT \cite{wang2017load}, a conventional network-centric LB method; and iii) SSS, a straightforward AP selection method without the capability of LB. The simulations are carried out in Python3.8 on a desk computer with an Intel Core i5-10500@3.1GHz processor, with the parameters summarised in Table \ref{Table: Parameters}. The relevant codes are open-sourced in GitHub \cite{MS-ATCNN-github}.

\begin{table}[t]
\renewcommand{\arraystretch}{1.2}
\centering
\caption{Parameter Setup}
\footnotesize
\begin{tabular}{c!{\vrule width 0.8pt}c} 
\Xhline{1pt}   
\textbf{HLWNet Parameters}     & \textbf{Values}  \\ 
\Xhline{1pt}  
Room size, $L \times W \times H$    & 10m$\times$10m$\times$3m    \\
Number of LiFi APs  & 16  \\
Number of WiFi APs  & 1  \\
LiFi AP separation     & 2.5m   \\
Height of WiFi AP    & 0.5m   \\
Number of UEs  & $[10, 100]$  \\
Average data rate requirement & 100 Mbps  \\
Other LiFi and WiFi parameters    & Refer to \cite{ji2023adaptive}  \\
\Xhline{1pt}
\textbf{Dataset Collection Parameters}         & \textbf{Values}  \\  
\Xhline{1pt}  
Throughput-degradation percentage, $\Delta\Gamma$ & 5\%  \\
Mobility model & RWP \\
Range of movement direction & $[0, 2\pi]$\\
Range of speed  & \ $(0,10]$ m/s \\
Sample number per AP type, $N$   & 2000   \\
Duration per sample, $T$  & 10s   \\
\Xhline{1pt}  
\textbf{Training Parameters}           & \textbf{Values}  \\ 
\Xhline{1pt}
Number of hidden neurons, $N_{L1}$ & 16   \\
Number of hidden neurons, $N_{L2}$ & 4   \\
Loss function    & MSE   \\
Learning rate, $\eta$  & 0.00001           \\
Optimiser    & Adam     \\
\Xhline{1pt}    
\end{tabular}\label{Table: Parameters}
\end{table}

\subsection{MS-ATCNN versus ATCNN}

In this subsection, the throughput performance of MS-ATCNN is compared with ATCNN to verify the effectiveness of the proposed mechanism, i.e., adaptive update interval. To guarantee a fair comparison, we let ATCNN adopt the same amount of average update interval as MS-ATCNN, of which the measured results are given in Table \ref{Table: Average update intervals}. This case is referred to as ATCNN@Aver. Two other cases of ATCNN are also considered: i) with a certain fixed update interval (e.g., ATCNN@10ms), ii) using a linear regression (LR) model to yield the update interval in accordance to the UE's speed\footnote{Two reference points in the format (speed, update interval) are adopted for the LR model: (1 m/s, 2s) and (10 m/s, 0.01s).}. The later case is referred to as ATCNN@LR. The impact of algorithm runtime is not involved here but will be analysed in the following subsections.

\begin{table}[t]
\centering
\caption{Average Update Interval of MS-ATCNN.}
\renewcommand{\arraystretch}{1.2}
\begin{tabular}{|c|c|c|c|c|c|}
\hline
Average speed (m/s) & 1 & 2 & 3 & 4 & 5 \\\hline
Average update interval (ms) & 1014 & 742 & 626 & 490 & 443 \\\hline
\end{tabular}\label{Table: Average update intervals}
\end{table}

\begin{figure}[t]
\centering
\includegraphics[height=3in,width=3.4in]{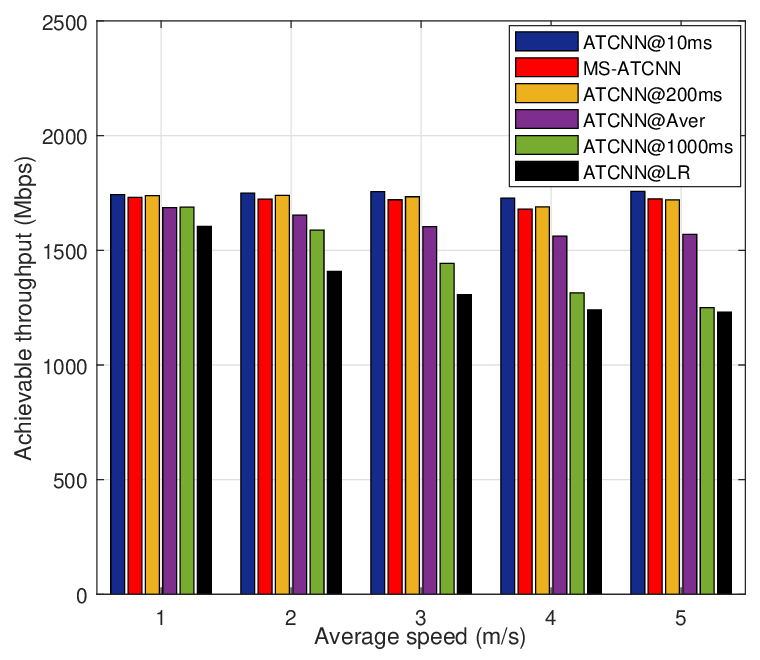}
\caption{Throughput comparison between MS-ATCNN and ATCNN with different update schemes (${N_u}=50$).}
\label{Fig: Thr_versus_UTI}
\end{figure}

Fig. \ref{Fig: Thr_versus_UTI} presents the overall network throughput achieved by MS-ATCNN, against ATCNN with different update schemes that are described above. Among all the schemes, ATCNN@10ms exhibits the highest throughput, as expected. Due to the lag effect, the throughput performance of ATCNN drops as the update interval increases, especially in the case of a higher UE speed. Also, it is observed that MS-ATCNN achieves a throughput comparable to ATCNN@200ms, while the average update interval of MS-ATCNN is much longer than 200 ms, resulting in a lower feedback cost. Compared to ATCNN@Aver, which employs the same amount of average update interval as MS-ATCNN, the later gains a noticeable throughput increase. Taken 5 m/s as an example, MS-ATCNN obtains a throughput of 1,740 Mbps, which is 10.1\% higher than the 1,580 Mbps achieved by ATCNN@Aver. Further, it proves that MSNN is more effective than the LR model, which is an intuitive adaptive update scheme. As can be seen, the throughput gap between MS-ATCNN and ATCNN@LR increases with the UEs' average speed, from 8.0\% at 1 m/s to 39.2\% at 5 m/s.

\subsection{MS-ATCNN versus Network-Centric Load Balancing}

\begin{figure}[t]
\centering
\includegraphics[height=3in,width=3.4in]{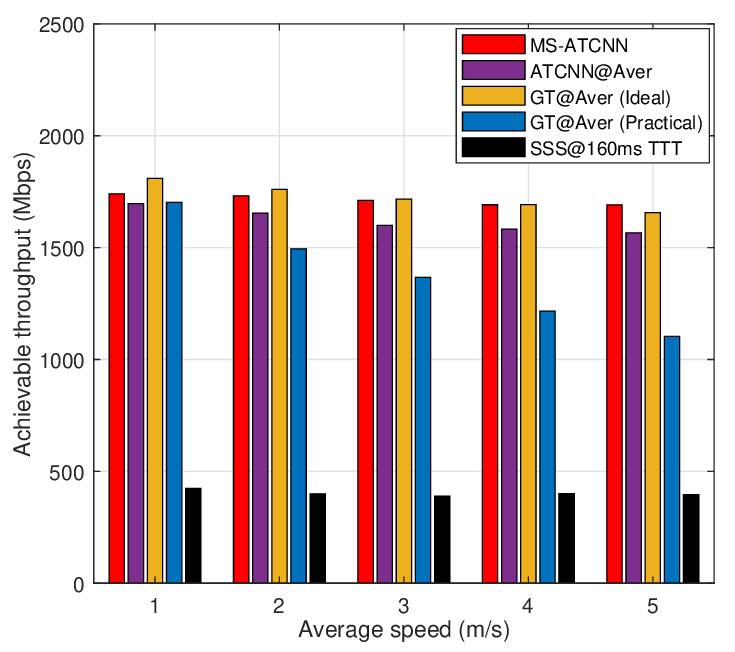}
\caption{Network throughput versus the UEs' average speed ($N_u$ = 50).}
\label{Fig: Thr_versus_Velocity}
\end{figure}

In this subsection, we focus on comparing the throughput performance of MS-ATCNN with GT, which is a typical network-centric LB method. Similar to ATCNN@Aver, GT uses the same amount of average update inverval as MS-ATCNN, in order to ensure a fair comparison. Accordingly, this method is referred to as GT@Aver. It is worth noting that GT requires a substantial amount of runtime, which would cause a lag effect and degrade the throughput performance. Therefore, two cases of GT are considered: the ideal case and the practical case, depending on whether the impact of runtime is taken into account. In contrast to GT, MS-ATCNN costs an ultra-low runtime, which is negligible in relation to the channel coherence time. Thus, the throughput results of MS-ATCNN with and without the impact of runtime are physically the same. For this reason, only the results of MS-ATCNN with the impact of runtime are presented here. In addition to GT, SSS in association with a standard handover process (which has a TTT of 160 ms \cite{3gpp_r13}) is also involved. This approach offers a baseline from two perspective: i) SSS itself exhibits the network capacity without the capability of LB, providing a comparison with MS-ATCNN in terms of resource management; and ii) the user-centric handover scheme renders a comparison with MS-ATCNN in regards to mobility management.

\subsubsection{Impact of the UEs' Average Speed} Fig. \ref{Fig: Thr_versus_Velocity} presents the achievable throughput of MS-ATCNN against the above baseline methods for different values of the UEs' average speed. As shown, the achievable throughput of MS-ATCNN is comparable to that of GT@Aver (Ideal). However, when involving the practical algorithm runtime, MS-ATCNN can significantly outperform GT@Aver (Practical), especially in a highly mobile environment. It is also found that the network throughput of MS-ATCNN reduces very slightly as the UEs move faster. When the UEs' average speed increases from 1 m/s to 5 m/s, the throughput obtained by MS-ATCNN reduces from 1,750 Mbps to 1,690 Mbps. In contrast, GT@Aver (Practical) exhibits a prominent decrease in throughput from 1,700 Mbps to 1,100 Mbps. Accordingly, the throughput gain achieved by MS-ATCNN against GT@Aver (Practical) enlarges from 2.9\% to 59.1\%. The reason behind this trend is that MS-ATCNN costs a much lower amount of runtime than GT. See a detailed analysis of the algorithm runtime in Section \ref{subsection:complexity}. Despite with user-centric handover, the throughput performance of SSS is far worse than the others, due to the lack of LB capability. Taken 5 m/s as an example, MS-ATCNN acquires a network throughout 322.5\% higher than SSS.

\begin{figure}[t]
\centering
\includegraphics[height=3in,width=3.4in]{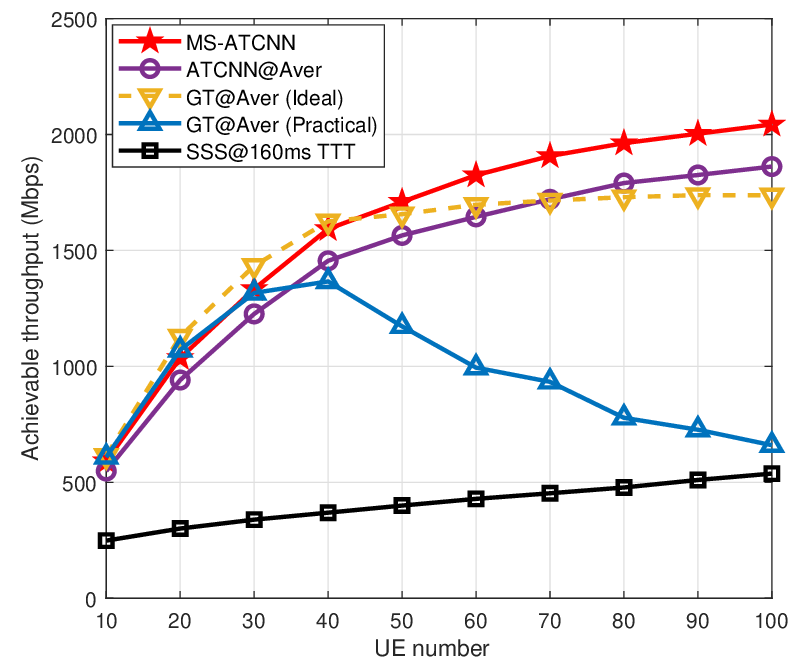}
\caption{Network throughput versus the number of UEs ($\bar{v}$ = 5 m/s).}
\label{Fig: Thr_versus_UEnum}
\end{figure}

\subsubsection{Impact of the Number of UEs} Fig. \ref{Fig: Thr_versus_UEnum} presents the network throughput as a function of the number of UEs ${N_u}$, with the UEs' average speed $\bar{v}$ set to be 5 m/s as an example. Three outcomes are observed. First, the achievable throughput increases with ${N_u}$ for all the involved methods, except GT@Aver (Practical). This is because the runtime of GT drastically increases with ${N_u}$, which will be discussed in the following subsection. Accordingly, the throughput gap between GT@Aver (Ideal) and GT@Aver (Practical) enlarges from 0\% to 62.2\% when ${N_u}$ increases from 10 to 100. Second, the network throughput of MS-ATCNN slightly falls behind that of GT@Aver (Ideal) when ${N_u}\leq$ 45, but the situation is opposite for a larger ${N_u}$. The reason for this trend is two-fold. On the one hand, the ATCNN model is trained with the dataset collected from GT, leading to an inevitable throughput gap, as discussed in \cite{ji2023adaptive}. On the other hand, MS-ATCNN benefits from its unique adaptive update scheme, in comparison to the fixed update scheme for GT. For a smaller ${N_u}$, the former factor dominates and hence the throughput of MS-ATCNN is marginally lower than that of GT@Aver (Ideal). While ${N_u}$ becomes larger, the latter factor dominates and thus MS-ATCNN surpasses GT@Aver (Ideal). Third, as ${N_u}$ increases from 10 to 100, MS-ATCNN achieves almost the same throughput as GT@Aver (Practical) at first until ${N_u}$ reaches 30, but afterwards MS-ATCNN outperforms GT@Aver drastically, resulting in a gain of 215\% when ${N_u}=$ 100. This trend is a combined effect of the previous two outcomes.

\subsubsection{Discussion}
The above observations signify that the proposed MS-ATCNN gains benefit over the existing LB methods from both perspectives of resource management and mobility management. In terms of resource management, ATCNN@Aver can achieve a higher throughput than GT@Aver (Practical), since ATCNN requires a much lower runtime than GT. As for mobility management, MS-ATCNN outperforms ATCNN@Aver due to the capability of adaptive update interval offered by MSNN. Combining the advantages of ATCNN and MSNN, MS-ATCNN can significantly outmatch the conventional network-centric LB methods.

\begin{figure}[t]
\centering
\subfigure[MSNN versus RNN on update interval accuracy ($N_u = 50$).]{\includegraphics[width=3.0in]{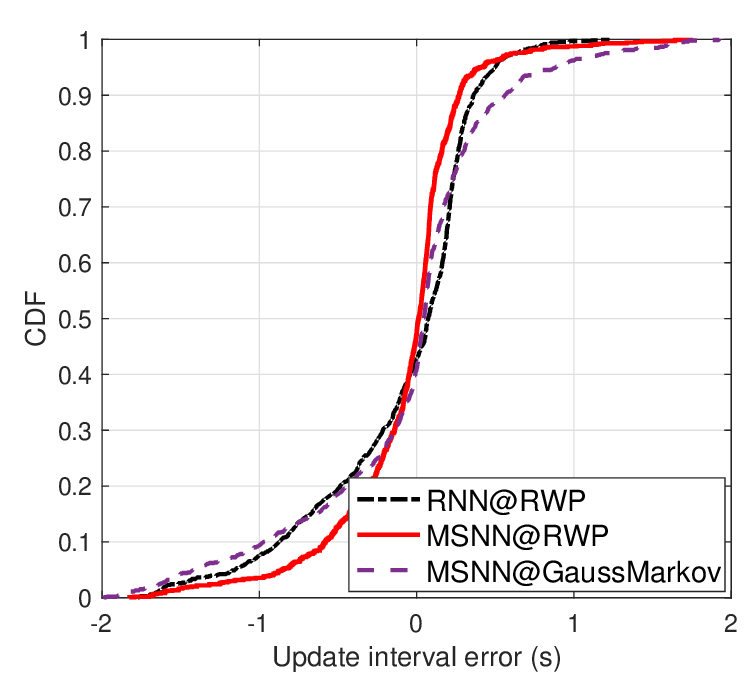}
\label{fig:CDF_MSNN_RNN}}
\subfigure[MS-ATCNN versus RNN-ATCNN on network throughput.]{\includegraphics[width=3.0in]{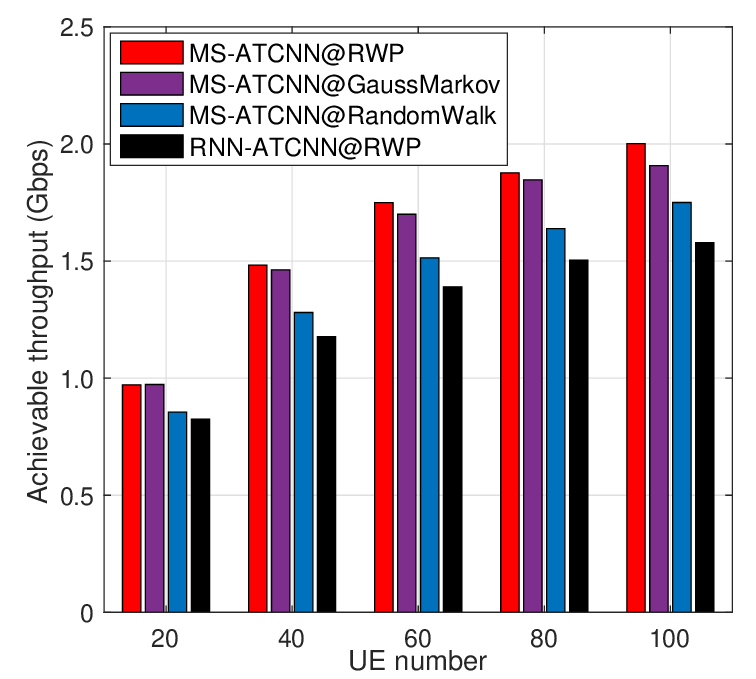}
\label{fig:Thr_MSNN_RNN}}
\caption{Robustness of MS-ATCNN against mobility models ($\bar{v}=2$ m/s).}
\label{Fig:MSNN_RNN}
\end{figure}

\subsection{Impact of Mobility Models}\label{subsection:robustness}
The above subsections examine the performance of MS-ATCNN that is trained and tested with the RWP model. To study the robustness of MS-ATCNN against other mobility models, the MS-ATCNN model trained with RWP is tested with the Gauss-Markov and random walk mobility models\footnote{The codes of the Gauss-Markov and random walk mobility models can be found in \url{https://github.com/panisson/pymobility}.}. The direction randomness and speed variance in the Gauss-Markov model are set to be 0.8 and 1. The flight length in the random walk model is fixed to be 20m. Considering the existing temporal data features in a mobile environment, the recurrent neural network (RNN)\footnote{Unlike MSNN, the RNN model is infeasible for being AP-specific, because the dataset would lose the temporal continuity if it is separated into different AP types. Instead, a unified RNN model is adopted for all APs.} is also considered as a baseline, of which the network structure and parameter setup refer to \cite{MS-ATCNN-github}.

As presented in Fig. \ref{fig:CDF_MSNN_RNN}, MSNN@RWP outperforms MSNN@Gauss-Markov in terms of estimating the desired update interval, since the MSNN model is specially trained with RWP. However, the gap between MSNN@RWP and MSNN@GaussMarkov is marginal, 
proving the robustness of the MSNN model against different mobility models. Apart from that, it is observed that RNN@RWP performs slightly worse than MSNN@RWP. The reason is two-fold. On the one hand, MSNN owns a competent capability of prediction in the case of RWP, due to the involvement of moving status. On the other hand, the AP-specific design of MSNN can effectively improve the estimation accuracy, while RNN fails to do so in nature.

In Fig. \ref{fig:Thr_MSNN_RNN}, the network throughput is compared between MS-ATCNN and RNN-ATCNN. As can be seen, MS-ATCNN@Gauss-Markov acquires a network throughput slightly lower than MS-ATCNN@RWP, with a gap less than 5\%. The reason has been explained above. Meanwhile, MS-ATCNN@RandomWalk is noticeably inferior to MS-ATCNN@Gauss-Markov. This is because the UE with the Gauss-Markov model shows a higher tolerance of update frequency errors than with random walk. Despite this, the throughput gap between MS-ATCNN@RandomWalk and MS-ATCNN@RWP is within 12\%. This verifies that MS-ATCNN is robust against different mobility models. In addition, it is found that RNN-ATCNN@RWP acquires a throughput up to 20\% lower than MS-ATCNN@RWP, in line with the outcome in Fig. \ref{fig:CDF_MSNN_RNN}.

\subsection{Analysis of Computational Complexity and Scalability}\label{subsection:complexity}

\begin{table}
\renewcommand{\arraystretch}{1.5}
\setlength\tabcolsep{1.5 mm}
\centering
\caption{Analysis of Big-O Complexity.}
\begin{tabular}{|c|c|c|}
\hline
\textbf{Methods} & \multicolumn{2}{c|}{\textbf{Big-O Complexity}}  \\ 
\hline
SSS  & \multicolumn{2}{c|}{$\mathcal{O}({N_{\rm{a}}}N_{\rm{u}})$}  \\ 
\hline
GT \cite{wang2017load} & \multicolumn{2}{c|}{$\mathcal{O}({N_{\rm{a}}}N_{\rm{u}}{I_{\rm{GT}}})$}  \\ 
\hline
\multicolumn{1}{|c|}{\multirow{2}{*}{MS-ATCNN}} & ATCNN \cite{ji2023adaptive}&  $\mathcal{O}({N_{\rm{a}}}MK_{m} + K_{o})$   \\ 
\cline{2-3}
\multicolumn{1}{|c|}{} & MSNN  &  $\mathcal{O}({N_{L_1}}{N_{L_2}})$  \\  
\cline{2-3}
\hline
\end{tabular}\label{Table: Computational Complexity}
\end{table} 

The Big-O complexity of MS-ATCNN is given in Table \ref{Table: Computational Complexity}, in comparison with SSS and GT. Here, $I_{\rm{GT}}$ is the number of iterations required by GT; $M$ denotes the maximum number of UEs that ATCNN can support ($M=$ 100 in this paper); $\mathcal{O}(K_{m})$ is the complexity of additions and multiplications in the FC layers; and $\mathcal{O}(K_{o})$ stands for the complexity of other operations including the BN and activation functions.
As reported by \cite{ji2023adaptive}, the complexity of ATCNN increases very slightly with $N_u$ and $N_a$, indicating that ATCNN has a strong scalability. As for MSNN, its complexity only relies on the dimension of the neural network structure (i.e., the product of ${N_{L_1}}$ and ${N_{L_2}}$). Hence, the complexity of MSNN remains the same when $N_a$ or $N_u$ increases. In summary, the proposed MS-ATCNN algorithm exhibits an excellent scalability in terms of the computational complexity.

\begin{figure}[t]
\centering
\includegraphics[height=2.8in,width=3.2in]{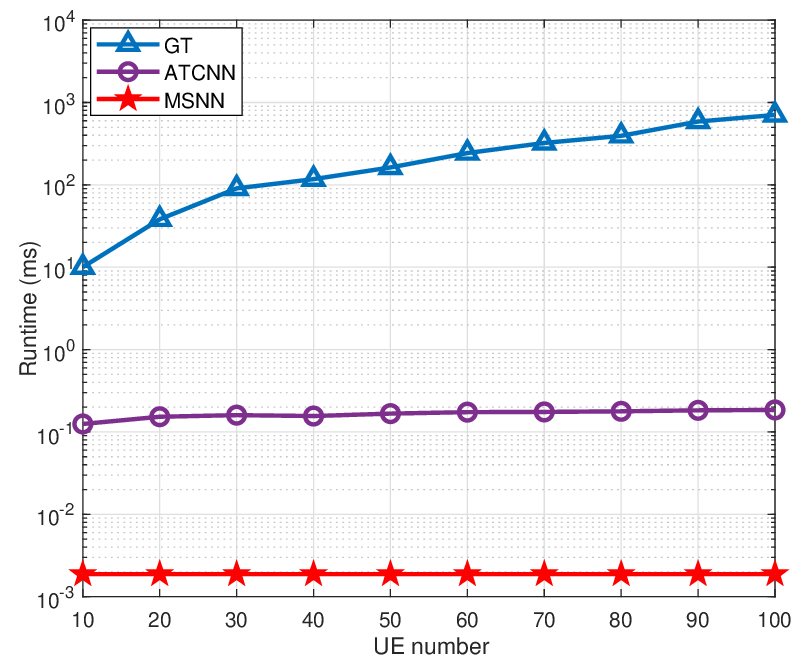}
\caption{Algorithm runtime versus the number of UEs.}
\label{Fig: Runtime_Plot}
\end{figure}

Fig. \ref{Fig: Runtime_Plot} shows the runtime as a function of the number of UEs for MSNN, ATCNN and GT. All the algorithms are run on MATLAB R2022b with the same configuration to ensure a fair comparison. It is observed that MSNN has a runtime two orders of magnitude lower than ATCNN. Specifically, it takes around 2 $\mu$s to implement MSNN, regardless of $N_u$. This fits the analysis of Big-O complexity for MSNN. Regarding ATCNN, its runtime increases from 120 $\mu$s to 184 $\mu$s when $N_u$ grows from 10 to 100. As a result, ATCNN plays a dominant role in counting the overall runtime of MS-ATCNN. In contrast, GT consumes a significant amount of runtime, which exhibits a prominent increase from 10 ms to 700 ms as $N_u$ increases from 10 to 100. The runtime of GT is about 80 times higher than that of MS-ATCNN when $N_u = 10$, and this gap soars to 3,800 times when $N_u = 100$.

\section{Conclusion} \label{sec:conclusion}
In this paper, a novel user-centric learning method named MS-ATCNN was proposed to tackle the joint resource and mobility management issue for hybrid networks such as HLWNets. The proposed MS-ATCNN consists of two key components: i) the ATCNN model which makes the LB solution for a target UE, from the perspective of resource management; and ii) the MSNN model which decides when the ATCNN model is implemented the next time, from the perspective of mobility management. Unlike the conventional network-centric LB schemes which can only be updated for all the UEs at the same pace, MS-ATCNN enables adaptive update frequencies among the UEs in accordance to their moving status. This is attributed to the unique feature of user-centric LB that resides in MS-ATCNN. Results show that with the same amount of average update interval, MS-ATCNN can significantly improve the network throughput against the conventional LB methods such as GT, with an increase up to 215\%. Apart from that, MS-ATCNN only requires a runtime at the level of sub-millisecond, which is two to three orders of magnitude lower than GT. For these reasons, MS-ATCNN can facilitate highly efficient spectrum aggregation in hybrid networks, delivering great potential to meet the requirements of future wireless communications. Future work will involve software defined network simulations and laboratory experiments to investigate the practicability of MS-ATCNN.

\bibliographystyle{IEEEtran}
\bibliography{IEEEabrv, Reference}

\end{document}